\documentclass[journal]{IEEEtran}
\usepackage{soul}
\usepackage{cite}
\usepackage{amsmath,amsfonts,amsxtra,amssymb,latexsym,amscd,amsthm,mathrsfs,bm}
\usepackage{algorithmic}
\usepackage{algorithm}
\usepackage{graphics}
\usepackage{graphicx}
\usepackage{hyperref}
\usepackage{textcomp}
\usepackage{balance}
\usepackage{booktabs}
\usepackage{caption}
\usepackage{subcaption}
\usepackage{cuted, nccmath}
\usepackage{amsmath}
\usepackage{lipsum}
\usepackage[dvipsnames]{xcolor}
\usepackage[utf8]{inputenc}
\usepackage[english]{babel}
\usepackage{filecontents}
\usepackage{xcolor}
\usepackage{thmtools}
\usepackage{multirow}

\usepackage{stackengine}
\usepackage{stfloats}

\usepackage{makecell}

\usepackage{caption}
\usepackage[compact]{titlesec}  
\titlespacing{\subsection}{0pt}{0.25pt}{0.25pt}
\titlespacing{\subsubsection}{0pt}{0.2pt}{0.2pt}
\makeatletter
\renewcommand{\fnum@figure}{Fig. \thefigure}
\makeatother

\DeclareMathOperator{\E}{\mathbb{E}}

\newcommand{\EX}[1]{\E\left\{{#1}\right\}}

\newcommand{\CG}[2]{\mathcal{CN}\left({#1},{#2}\right)}

\DeclareRobustCommand{\bigO}{%
  \text{\usefont{OMS}{cmsy}{m}{n}O}%
}

\newcommand{\dl}{\mathrm{d}}

\newtheoremstyle{mythm}%
{3pt}
{3pt}
{}
{}
{\bfseries}
{}
{.5em}
{}%

\declaretheoremstyle[
headpunct=\textup{:},
bodyfont=\normalfont,
]{myremark}
\theoremstyle{myremark}
\newtheorem{remarkex}{Remark}

\newenvironment{remark}
{\pushQED{\qed}\remarkex}
{\popQED\endremarkex}

\ifCLASSINFOpdf
\else
\fi
\begin{document}
%
\title{\huge Joint Access Point Activation and Power Allocation for Cell-Free Massive MIMO Aided ISAC Systems}
%
%

\author{Nguyen~Xuan~Tung, Le~Tung~Giang, Trinh~Van~Chien, Hoang~Trong~Minh, Lajos~Hanzo, ~\IEEEmembership{Life Fellow,~IEEE}
    \thanks{L. Hanzo would like to acknowledge the financial support of the Engineering and Physical Sciences Research Council (EPSRC) projects under grant EP/Y026721/1, EP/W032635/1 and EP/X04047X/1 as well as of the European Research Council's Advanced Fellow Grant QuantCom (Grant No. 789028). \textit{(Corresponding author: Nguyen Xuan Tung.)}}
    \thanks{Nguyen Xuan Tung is with Faculty of Interdisciplinary Digital Technology, PHENIKAA University, Yen Nghia, Ha Dong, Hanoi 12116, Viet Nam
    (e-mail: tung.nguyenxuan@phenikaa-uni.edu.vn). 
    Le Tung Giang is with the Department of Information Convergence Engineering, Pusan National University, Busan 46241, Republic of Korea (email: {giang.lt2399144@pusan.ac.kr}). 
    Trinh Van Chien is with the School of Information and Communication Technology (SoICT), Hanoi University of Science and Technology (HUST), Vietnam (email: chientv@soict.hust.edu.vn). 
    Trong-Minh Hoang is with the Posts and Telecommunications Institute of Technology, Vietnam (e-mail: hoangtrongminh@ptit.edu.vn). 
    Lajos Hanzo is with the Department of Electronics and Computer Science, University of Southampton, Southampton SO17 1BJ, U.K. (e-mail: lh@ecs.soton.ac.uk).}  
}

%
%

\markboth{}
{Shell \MakeLowercase{\textit{et al.}}: Joint Access Point Activation and Power Allocation for Cell-Free Massive MIMO Aided ISAC Systems}
%



\maketitle


\begin{abstract}
    \textcolor{black}{Cell-free massive multiple-input multiple-output (MIMO)-aided integrated sensing and communication (ISAC) systems are investigated where distributed access points jointly serve users and sensing targets. We demonstrate that only a subset of access points (APs) has to be activated for both tasks, while deactivating redundant APs is essential for power savings.} This motivates joint active AP selection and power control for optimizing energy efficiency. The resultant problem is a mixed-integer nonlinear program (MINLP). To address this, we propose a model-based Branch-and-Bound approach as a strong baseline to guide a semi-supervised heterogeneous graph neural network (HetGNN) for selecting the best active APs and the power allocation. 
    \textcolor{black}{Comprehensive numerical results demonstrate that the proposed HetGNN reduces power consumption by $\mathbf{20-25\%}$ and runs nearly $\mathbf{10,000}$ times faster than model-based benchmarks.}

\end{abstract} 
\IEEEpeerreviewmaketitle
\section{Introduction}

Integrated sensing and communication (ISAC) services are vital for next-generation wireless networks, enabling spectrum and hardware reuse at a modest added cost \cite{OpenIssue_JSAC_Lu_Feb_2024}. Sensing in ISAC also enhances communication via improved localization, resource allocation, and CSI tracking. For fully harnessing these benefits, numerous studies have been conducted to optimize beamforming \cite{beamforming_TVT_2023}, radar signal processing \cite{waveform_TSP_2016}, and resource allocation \cite{ISAC_JSAC_2022}. Early ISAC research primarily aimed for single integrated dual-function at transmitters, categorized into single-antenna systems leveraging waveform diversity \cite{waveform_TSP_2016} and multi-antenna systems optimizing beamforming \cite{beamforming_TVT_2023}. Recent research has then also evolved to multi-transmitter ISAC systems \cite{TWC_ISAC_Behdad_Sep_2024, TVT_ISAC_Huang_Jul_2022}, where distributed transmitters and receivers collaborate under a Cloud Radio Access Network (C-RAN) relying on centralized management. 

An emerging topic is that of integrating sensing into cell-free massive MIMO (cf-mMIMO) systems to leverage its distributed access points (AP) architecture for enhancing coverage, spatial diversity, and target detection accuracy \cite{Mao10516289}. However, optimizing ISAC in cf-mMIMO systems presents challenges due to the escalating system complexity. Conventional methods decouple radar sensing and communication, leading to suboptimal resource utilization and scalability issues \cite{TVT_ISAC_Huang_Jul_2022}. Moreover, the computational overhead becomes prohibitive in large-scale networks. 
\textcolor{black}{Recent AI‐driven ISAC studies confirm that deep learning (DL)-based approaches are capable of addressing these issues, significantly improving both sensing accuracy and resource allocation efficiency \cite{2024_ComMag_Wu_ISAC_AI}. However, they often lack generalization to scenarios beyond their training configuration.} 
Graph neural networks (GNNs) have emerged as a promising solution, modeling complex relationships and ensuring scalability by generalizing across network configurations \cite{Tung10750215, CentraliedGNN_Yifei2022, 2024_Tung_GNNSurvey}. By representing a wireless system as a graph, where nodes and edges capture system entities and interactions, GNNs succeed in predicting outcomes based on node and edge features rather than fixed system dimensions. Hence, they are well-suited for dynamic wireless environments.

Against this backdrop, we propose a novel heterogeneous GNN (HetGNN) for resource allocation in a cell-free massive multiple-input multiple-output (MIMO)-based C-RAN, where dual-function APs serve users and perform sensing by cooperating with receiving stations (SRs) via the C-RAN controller. The goal is to minimize total power consumption while meeting the UE's data throughput and sensing accuracy requirement under the Cramer-Rao bound (CRB). We will demonstrate that activating all APs, as in \cite{Vu7029696}, leads to unnecessary service and resource wastage. Explicitly, utilizing only a necessary subset is more efficient, especially since the AP locations directly impact power allocation, hence affecting both communication and sensing. This motivates the joint optimization of AP selection and power control. To the best of our knowledge, this is the first study to jointly optimize both the AP activation and power allocation for both communication and sensing. Given the mixed-integer nonlinear (MINL) nature of the problem, we model the ISAC system as a heterogeneous graph and harness a HetGNN for learning both APs activation and power control strategies while ensuring constraint satisfaction. The semi-supervised HetGNN conceived is trained using guidance from a proposed model-based Branch-and-Bound approach. Numerical results demonstrate that the designed HetGNN outperforms optimization-based methods, achieving significantly reduced power consumption and prompt execution.

\emph{Notation:} Boldface lowercase letters and boldface uppercase letters denote vectors and matrices, respectively. Let $()^T$ and $()^*$ denote the transpose and conjugate, respectively. The Euclidean norm and expectation are denoted by $\|\cdot\|$, and $\mathbb{E}\{\cdot\}$. Finally, we define the circularly symmetric complex Gaussian distribution with variance $\sigma^2$ by $\mathcal{CN}(0,\sigma^2)$.

\section{Integrated Sensing and Downlink Cellfree-Massive MIMO Communication} \label{SystemModel}
\subsection{System and Signal Models}
\begin{figure}[t]
    \centering
    \includegraphics[width=0.8\linewidth]{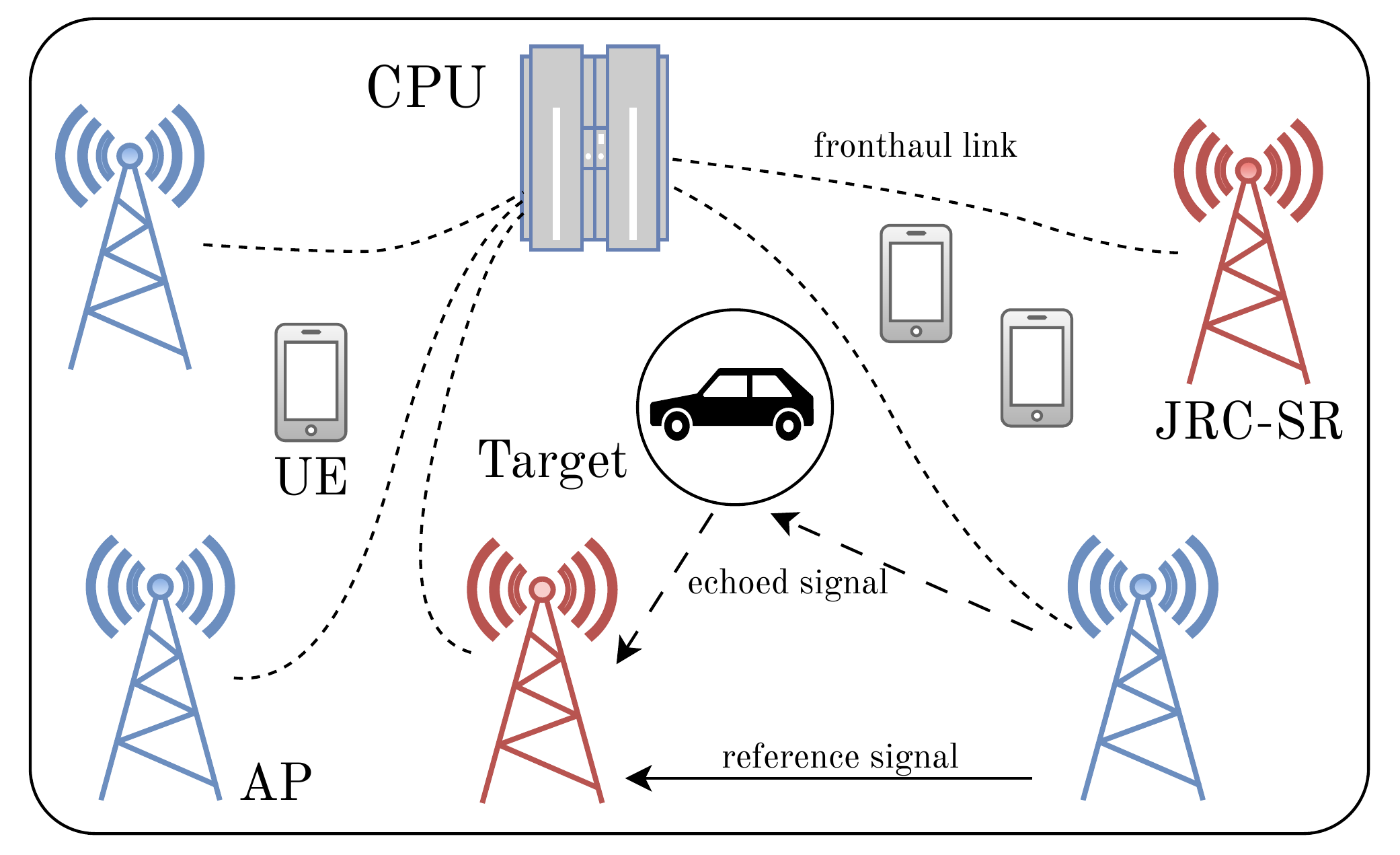}
    \caption{Illustration of the cf-mMIMO ISAC system.}
    \label{fig:system-model}
    \vspace{-0.75cm}
\end{figure}

We consider a cf-mMIMO ISAC system where a set of UEs, $\mathcal{K} = \{1,..., K\}$, are jointly served by a set of dual-function APs, $\mathcal{M} = \{1,..., M\}$, as illustrated in Fig.~\ref{fig:system-model}. Each AP sends a combined signal to all UEs in support of efficient resource utilization. {\textcolor{black}{The downlink cf-mMIMO communication signal is exploited for localizing a single-point target vehicle within the area, facilitating further use in future applications. The signals transmitted from all APs are reflected by the sensing target, e.g., a vehicle, and are collected by a set $\mathcal{T}=\{1,..., T\}$ of sensing receivers (SRs). Following the passive sensing philosophy, each SR features two signals: the target’s echoed signal and the reference signals received from APs. By comparing these signals, key target parameters such as the location, speed, and Angle of Arrival (AoA) are estimated \cite{Godrich2011}.}} For simplicity, we focus our attention on the accuracy of target localization. All the $M$ APs and $T$ SRs are connected to the C-RAN controller \cite{CellFree_vs_SmallCell_hien2017} for resource management and signal processing.  
\subsubsection{Communication model}
We first consider the communication between APs and UEs, in the uplink pilot training phase, when each AP estimates the propagation channels impinging upon all UEs. The minimum mean-square error estimate (MMSE) of the channel between the $k$-th AP and the $n$-th UE is given by
{\color{black}
\begin{equation}
    \begin{aligned}
        \hat{h}_{mk} &= \frac{\mathbb{E}\{\tilde{y}_{p,mk}^* h_{mk}\}}{\mathbb{E}\{|\tilde{y}_{p,mk}|^2\}} \tilde{y}_{p,mk} \\
        &= \frac{\sqrt{\tau_p \rho_p} \varsigma_{mk}}{\tau_p \rho_p \sum\limits_{k'\in \mathcal{K}} \varsigma_{mk'} |\boldsymbol{\theta}_k^H \boldsymbol{\theta}_{k'}|^2 + \sigma^2_{\text{UL}}}\tilde{y}_{p,mk}, 
    \end{aligned}
\end{equation}
where $\tilde{y}_{p,mk} = \boldsymbol{\theta}_n^H\boldsymbol{y}_{p,m}$, and $\boldsymbol{y}_{p,m} = \sqrt{\tau_p \rho_p} \sum\limits_{k\in \mathcal{K}} h_{mk} \boldsymbol{\theta}_k + \boldsymbol{\xi}_{p,m}$ is the pilot signal received at the $k$-th AP.
} Here, $h_{mk} \sim \mathcal{CN}(0,\varsigma_{mk})$ and $\varsigma_{mk}$ denote the channel coefficient and the large-scale fading between the $m$-th AP and the $k$-th UE. Meanwhile, $\tau_p$, $\rho_p$, and $\boldsymbol{\theta}_k$, and $\boldsymbol{\xi}_{p,m}$ are the uplink training duration, the transmit pilot power, the pilot sequence used by the $k$-th UE, and the additive noise at the $m$-th AP, respectively. The channel estimate is distributed as $\hat{h}_{mk}\sim \mathcal{CN}(0,v_{mk})$, in which $v_{mk}\triangleq \mathbb{E}\{|\hat{h}_{mk} |^2\} = \frac{{\tau_p \rho_p} \varsigma^2_{mk}}{\tau_p \rho_p \sum_{k'\in \mathcal{K}} \varsigma_{mk} |\boldsymbol{\theta}_k^H \boldsymbol{\theta}_{k'}|^2 + 1}$.

Note that the UEs are distributed heterogeneously, hence, their throughput directly depends on their downlink signal quality gleaned from nearby APs, rather than from distant APs. Thus, activating only a subset $\mathcal{M}_{\text{act}}$ of APs may suffice for satisfying the UEs' requirement, despite turning off the remaining APs for conserving energy. Given the channel estimate obtained from the pilot training phase, the $m$-th AP transmits the combined signal, $x_m = \sum\limits_{k\in\mathcal{K}} \sqrt{P_{mk}} \hat{h}^*_{mk} s_k $, to the UEs. Here, $s_k$ with $\EX{|s_k|^2}=1$ and $P_{mk}$ are the symbol intended for the $k$-th UE and the corresponding allocated power, respectively. 
Accordingly, the signal received at the $k$-th UE is given by
\begin{align}\label{eq:rk1}
y_{\dl,k} 
&= 
\sum_{m\in \mathcal{M}_{\text{act}}} h_{mk}s_{k} +  
\xi_{\dl,k}\nonumber\\
 &=
\sqrt{\rho_d}\sum_{m\in\mathcal{M}_{\text{act}}}\sum_{k'\in \mathcal{K}}
\sqrt{P_{mk'}}h_{mk}\hat{h}_{mk'}^* s_{k'} + \xi_{\dl,k},
\end{align}
where $P_{mk'}$ represents the downlink (DL) transmission power allocated by the $m$-th AP to transmit the symbol $s_{k'}$. Here,  $\xi_{\dl,k} \sim \CG{0}{1}$ is the additive noise at the $k$-th UE. Note that the desired signal $s_k$ is detected from $y_{\dl,k}$ in \eqref{eq:rk1}. \textcolor{black}{The closed-form expression of the Signal-to-Noise-Ratio (SINR) of the $k$-th user is adopted from \cite{CellFree_vs_SmallCell_hien2017}, as expressed in \eqref{eq:Theo_rateexpr1}, where $\sigma^2_{\mathrm{DL}}$ denotes the noise power normalized by the downlink transmit power $\rho_{\mathrm{DL}}$.}
\begin{figure*}[t]
\begin{align}\label{eq:Theo_rateexpr1}
\gamma_{k}
 =
 \frac{\left(\sum\limits_{m\in \mathcal{M}_{\text{act}}} \sqrt{P_{mk}} v_{mk} \right)^2 }{ \sum\limits_{k'\neq k, k' \in \mathcal{K}}\left(\sum\limits_{m\in \mathcal{M}_{\text{act}}} 
    \sqrt{P_{mk'}}v_{mk}{\varsigma_{mk}}/{\varsigma_{mk'}} \right)^2| \pmb{\theta}_{k'}^H\pmb{\theta}_{k}|^2 + \sum\limits_{k'\in \mathcal{K}}\sum\limits_{m\in \mathcal{M}_{\text{act}}} P_{mk'} v_{mk'}\varsigma_{mk} + \sigma^2_{\text{DL}} }.
\end{align}
\hrulefill
\vspace{-0.5cm}
\end{figure*}

\subsubsection{Sensing model}
In the sensing context, a cross-ambiguity function is calculated between the reference signal and the target echoed signal \cite{Malanowski2012}. Let APs and SRs be located in a 2-D coordinate system as $[ x^{\textrm{AP}}_{m}, y^{\textrm{AP}}_{m}]$ and $[ x^{\textrm{SR}}_{t}, y^{\textrm{SR}
}_{t}]$, respectively. The coordinates of the target are denoted by $[x,y]$.
Then, the distance from the $m$-th AP to the target is $R^{\textrm{AP}}_{m} =  \sqrt{(x^{\textrm{AP}}_{m} - x)^2 + (y^{\textrm{AP}}_{m} - y)^2}$. Similarly, the distance from the $t$-th SR to the target is $R^{\textrm{SR}}_{t} = \sqrt{(x^{\textrm{SR}}_{t} - x)^2 + (y^{\textrm{SR}}_{t} - y)^2}$. The propagation delay from the $m$-th AP to the $t$-th SR is $\Delta t_{mt} = \frac{ R^{\textrm{AP}}_{m} +  R^{\textrm{SR}}_{t}}{c}$, where $c=3\times 10^{8}\mbox{[m/s]}$ is the propagation velocity of the signal. At the time $l$, the signal received at the $m$-th SR is expressed as:
\begin{align}
    r_{t}(l) = \sum_{m\in\mathcal{M}} \chi_{mt} \sqrt{P_k}s_{k}(l - \Delta t_{mt}) + z_{t}(t_{0}),
\end{align}
where $z_{t}(l)\sim \mathcal{CN}(0, \sigma_{S}^2)$ is the AWGN at the $t$-th SR, the coefficient $\chi_{mt}$ represents the effect of the target radar cross section (RCS), which are independent from each other. 

For any unbiased (or asymptotically unbiased) estimator of unknown parameter(s), the CRB provides a lower bound on the mean-square error (MSE) of target location estimation. This bound is commonly referred to as the Cramér-Rao Lower Bound (CRLB) and is derived in \cite{Godrich2011}. Accordingly, CRLB's matrix representation for target location is defined as:
\begin{align}
    \textbf{C}_{x,y} (\bm{p}_{\text{sen}}) = \left\{ \sum_{m \in \mathcal{M}_{\textrm{act}}} P_{\text{sen},m} 
    \begin{bmatrix}
q_{a_m} & q_{c_m} \\
q_{c_m} & q_{b_m} 
\end{bmatrix} \right\}^{-1},
\end{align}
where $\bm{p}_{\text{sen}} = [P_{\text{sen},0},...,P_{\text{sen},m},...,P_{\text{sen},M}]^T$, with $P_{\text{sen},m} = \sum\limits_{k\in\mathcal{K}}P_{mk}$ and $P_{\text{sen},m} = 0, \forall m\not\in\mathcal{M}_{\textrm{act}}$. Furthermore, $q_{a_m}= \zeta \sum_{t\in\mathcal{T}} |\chi_{mt}|^2 \left(\frac{x^{\textrm{AP}}_{m} - x}{R^{\textrm{AP}}_{m}} + \frac{x^{\textrm{SR}}_{t} - x}{R^{\textrm{SR}}_{t}}\right)^2$, $q_{b_m} = \zeta \sum_{t\in\mathcal{T}} |\chi_{mt}|^2 \left(\frac{y^{\textrm{AP}}_{m} - y}{R^{\textrm{AP}}_{m}} + \frac{y^{\textrm{SR}}_{t} - y}{R^{\textrm{SR}}_{t}}\right)^2$, and $q_{c_m} = \zeta \sum_{t\in\mathcal{T}} |\chi_{mt}|^2 \left(\frac{x^{\textrm{AP}}_{m} - x}{R^{\textrm{AP}}_{m}} + \frac{x^{\textrm{SR}}_{t} - x}{R^{\textrm{SR}}_{t}}\right)\left(\frac{y^{\textrm{AP}}_{m} - y}{R^{\textrm{AP}}_{m}} + \frac{y^{\textrm{SR}}_{t} - y}{R^{\textrm{SR}}_{t}}\right)$, where $\zeta = \frac{8\pi^2 B^2}{\sigma_{S}^{2} c^2}$. The sum of CRLBs used for estimating target location is expressed as:

\begin{align}
    \sigma_{x,y}^2(\bm{p}_{\text{sen}}) = \frac{\bm{b}^{\textrm{T}} \bm{p}_{\text{sen}}}{\bm{p}_{\text{sen}}^{\textrm{T}} \bm{A} \bm{p}_{\text{sen}}}, 
\end{align}
where 
\begin{align}\label{eq:b_A}
\begin{aligned}
    \bm{b} &= \bm{q}_{a} + \bm{q}_{b}, \\
    \bm{A} &= \bm{q}_{a}\bm{q}_{b}^{\textrm{T}} - \bm{q}_{c}\bm{q}_{c}^{\textrm{T}},
\end{aligned}
\end{align}
with $\bm{q}_{a} = [g_{a_1}, \cdots,g_{a_M}]^{\textrm{T}}, \bm{q}_{b} = [g_{b_1}, \cdots,g_{b_M}]^{\textrm{T}}, \text{ and } \bm{q}_{c} = [g_{c_1}, \cdots,g_{c_M}]^{\textrm{T}}$. The target location is updated in real time to determine the localization accuracy at each time step.

\subsection{Problem Formulation}

Our objective is to minimize the total transmit power across all RRUs while simultaneously meeting both the communication throughput and the sensing accuracy requirements. The optimization problem is mathematically formulated as follows:
\begin{subequations}
\label{P2:problem}
    \begin{align}
        \underset{\bm{p}_{\text{sen}}, \mathcal{M}_{\text{act}}}{\textrm{minimization}} \quad\quad& \sum\limits_{m\in \mathcal{M}_{\text{act}}}\sum\limits_{k\in\mathcal{K}} P_{mk} + \sum\limits_{m\in \mathcal{M}_{\text{act}}}P^{\textrm{cir}} \label{P1:obj}\\
                                    \textrm{s.t} \qquad \qquad & \sum\limits_{k\in\mathcal{K}}v_{mk}P_{mk}\leq 1, \forall m\in\mathcal{M}, \label{P1: Power constraint}\\
                                    & \gamma_{k} \geq \gamma^{\textrm{thr}}, \forall k \in \mathcal{K}, \label{P2:gamma_thr}\\
                                    &  \bm{b}^{T} \bm{p}_{\text{sen}} -\nu \bm{p}_{\text{sen}}^T \bm{A} \bm{p}_{\text{sen}} \leq 0. \label{P2:CRLB} 
    \end{align}
\end{subequations}
\textcolor{black}{Here, $P^{\textrm{cir}}$ denotes the power required to activate and maintain a single AP for the cf-mMIMO DL communication.} Constraint \eqref{P1: Power constraint} limits the power budget, \eqref{P2:gamma_thr} ensures meeting the minimum SINR for UEs, and \eqref{P2:CRLB} enforces the CRLB-based sensing accuracy with threshold $\nu$.

With a fixed set of active APs, the power allocation problem becomes convex by relaxing constraint \eqref{P2:CRLB} and reformulating \eqref{P2:problem}, allowing efficient solutions via tools like CVX \cite{grant2015cvx}. However, determining the optimal set of active APs involves evaluating a total of $\sum\limits_{i=1}^M \frac{M!}{i! (M-i)!}$ configurations. This makes exhaustive search infeasible, particularly in large-scale networks. While heuristic approaches can accelerate the search process, they often yield suboptimal solutions. This inspires the search for a more efficient solution having improved performance, yet reduced complexity.

\section{Heterogeneous GNN-based Approach}
In this section, we introduce a novel learning-based framework, referred to as the HetGNN-based model, to address the power allocation problem in \eqref{P2:problem}. Specifically, we first propose a modified Branch-and-Bound (modified-BB) technique for determining a set of active APs and simultaneously allocate power. The modified-BB approach generates labels to guide the semi-supervised training of the proposed HetGNN model, using a small weighting coefficient. 

\subsection{Modified-BB Algorithm}\label{sec: modified BB}

\begin{algorithm}[t]
    \caption{AP Activation and Power Allocation via Branch-and-Bound with Relaxed Constraints} 
    \label{BB Algorithm} 
    \begin{algorithmic}[1] 
        \STATE \textbf{Input}: System parameters $\mathcal{M}$, $\mathcal{K}$, and $\mathcal{T}$.
        \STATE Initialize the BB tree with all APs active, $\mathcal{M}_{\text{act}} = \mathcal{M}$.
        \STATE Solve \eqref{P3:relaxed_problem} and calculate $P_{\textrm{tot}}^{\textrm{curr}}= \sum\limits_{m\in \mathcal{M}_{\text{act}}}\sum\limits_{k\in\mathcal{K}} P_{mk}$. Set the level counter $i=0$.
        \REPEAT 
            \STATE $i = i + 1$.
                \FOR {each AP $m' \in \mathcal{M}_{\text{act}}$} 
                    \STATE Set $\mathcal{M}^{\text{temp}}_{\text{act}} = \mathcal{M}_{\text{act}} \setminus \{m'\}$.
                    \STATE Solve the power allocation problem in \eqref{P3:relaxed_problem} for $\mathcal{M}^{\text{temp}}_{\text{act}}$. Calculate $P_{\textrm{tot}}^{\textrm{temp}}= \sum\limits_{m\in \mathcal{M}_{\text{act}}}\sum\limits_{k\in\mathcal{K}} P_{mk}$.
                    \IF {Problem is infeasible}
                        \STATE Prune the branch. 
                    \ELSE
                        \IF {$P_{\textrm{tot}}^{\textrm{temp}}\leq P_{\textrm{tot}}^{\textrm{curr}}$}
                        \STATE Retain the branch for further exploration.
                        \ENDIF
                    \ENDIF
                \ENDFOR
        \UNTIL {No nodes are pruned.} 
        \STATE {\textbf{Output}:Best solution found.}
    \end{algorithmic}
\end{algorithm} 

We construct a Branch-and-Bound (BB) tree that begins with all APs activated. We assume that the problem in \eqref{P2:problem} is feasible. At the first level of the BB tree, the system iteratively removes an AP $k$, forming a new active AP set $\mathcal{M}^{\text{temp}}_{\textrm{act}} = \mathcal{M} \setminus \{k\}$. The problem in \eqref{P2:problem} is then solved for this updated set $\mathcal{M}^{\text{temp}}_{\text{act}}$. 
\textcolor{black}{
To find solution for problem \eqref{P2:problem}, following \cite{Godrich2011}, we first relax the non‐convex constraint \eqref{P2:CRLB} as $\bm{b} -\nu \bm{A} \bm{p}_{\text{sen}} \leq 0$. 
This relaxation results in a convex CRLB constraint and guarantees that the minimum transmit power is close or equal to the true optimum of problem \eqref{P2:problem}.}
Furthermore, we introduce an auxiliary variable. Then we can reformulate the problem in \eqref{P2:problem} as
\begin{subequations}
\label{P3:relaxed_problem}
    \begin{align}
        \underset{\bm{p}_{\text{sen}}, \{y_{mk}\}}{\textrm{minimization}} \quad\quad& \sum\limits_{m\in \mathcal{M}^{\text{temp}}_{\textrm{act}}}\sum\limits_{k\in\mathcal{K}} y^2_{mk} \label{P3:obj}\\
        \textrm{s.t} \qquad \qquad & P_{mk}\geq y^2_{mk}, \forall m,k\\
        & \eqref{P1: Power constraint}, \eqref{eq: constraint rate relaxed},\\
        &  \bm{b} -\nu \bm{A} \bm{p}_{\text{sen}} \leq 0. \label{P3:CRLB} 
    \end{align}
\end{subequations}
The constraint in \eqref{eq: constraint rate relaxed} is reformulated from \eqref{P2:gamma_thr}, as illustrated at the top of the page. The optimization problem in \eqref{P3:relaxed_problem} is now in convex form and can be efficiently solved using the CVX toolbox \cite{grant2015cvx}. If the problem is infeasible, the branch is pruned, as the corresponding active AP set is not viable. Otherwise, the set $\mathcal{M}^{\text{temp}}_{\text{act}}$ is kept, if the total power consumption of the temporary structure is smaller than the previous structure. The process continues recursively, removing and evaluating APs at each level until no further APs can be removed without violating feasibility constraints. The detailed procedure of the modified BB technique is summarized in Algorithm \ref{BB Algorithm}. 

\begin{figure*}[t]
\begin{align}\label{eq: constraint rate relaxed}
    \sqrt{\gamma^{\textrm{thr}}} \sqrt{\sum_{k'\neq k, k'\in \mathcal{K}} \left(\sum_{m\in \mathcal{M}} y_{mk'} v_{mk} \frac{\varsigma_{mk}}{\varsigma_{mk'}} \right)^2 + \sum_{k'\in \mathcal{K}}\sum_{m\in \mathcal{M}}y_{mk'}^2 v_{mk'}\varsigma_{mk} + \sigma^2_{\text{DL}}} \leq \sum_{m\in \mathcal{M}} y_{mk} v_{mk}, \forall k\in\mathcal{K}
\end{align}
\hrulefill
\vspace{-0.5cm}
\end{figure*}
\begin{remark} \label{rmk:complexity}
By setting the condition in the $13$-th step, the modified BB algorithm prunes the BB tree, retaining only a suffice branch per level. This reduces the search space and lowers its complexity to \( \mathcal{O}(M^2) \). This is a clear improvement compared to the case keeping all possible branches associated with the complexity of \( \mathcal{O}(M^3) \). This reduction is crucial since the inner loop solves \eqref{P3:relaxed_problem} having \( 2MK \) optimization variables and \( MK + M + K \) constraints using CVX in each iteration, resulting in a complexity of \( \mathcal{O}[(2MK)^3] \). The modified approach significantly reduces the overall computational burden.


\end{remark}
\subsection{HetGNN-based Model}
\begin{figure}[t]
     \centering
    \includegraphics[width=0.85\columnwidth]{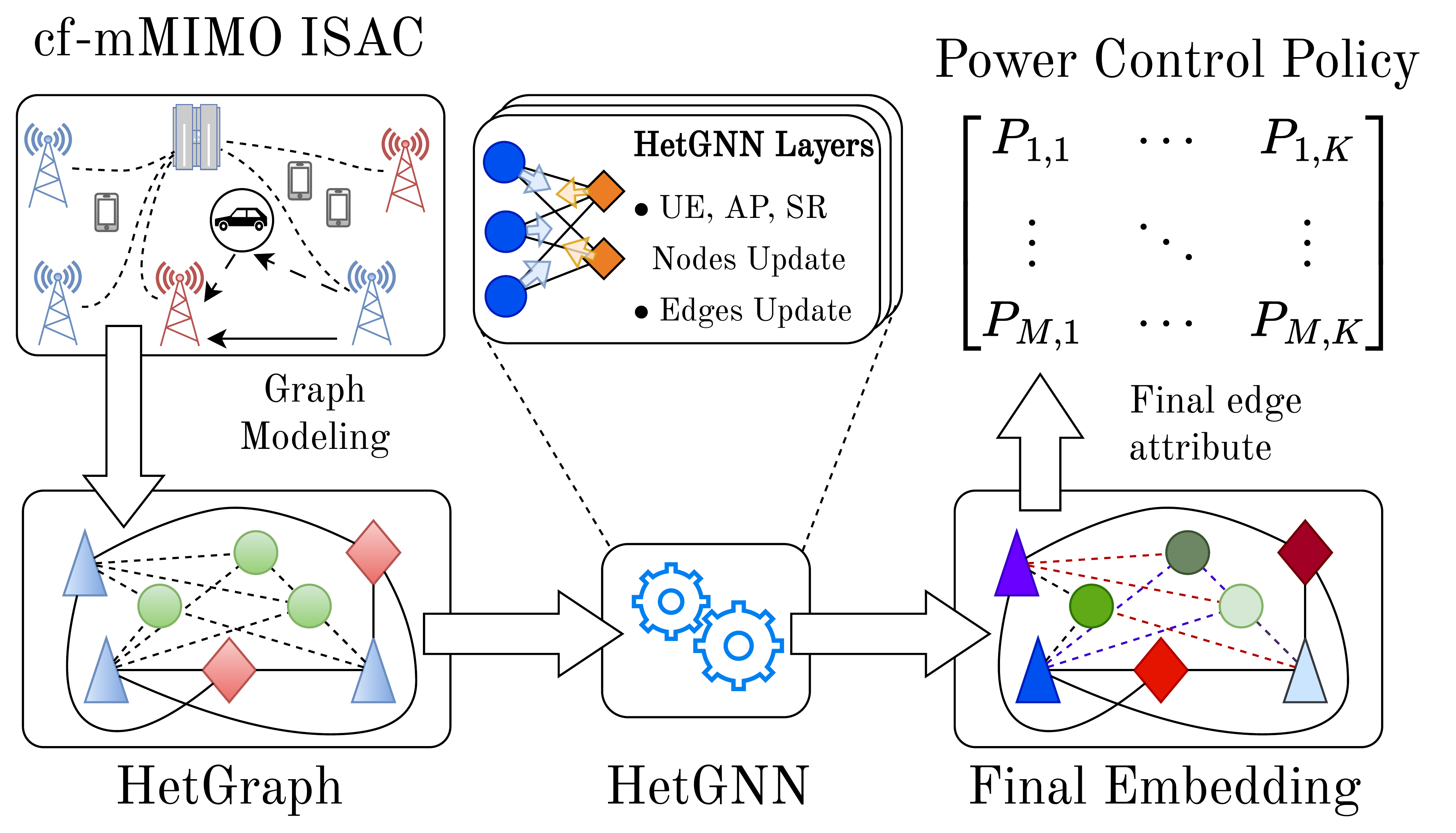}
    \caption{The general flow chart of the training phase.}
    \label{fig:HetGNN}
    \vspace{-0.5cm}
\end{figure}

The proposed HetGNN-based model consists of three main steps: (i) graph modeling, (ii) graph convolutional layer, and (iii) downstream task, as illustrated in Fig~\ref{fig:HetGNN}.

\subsubsection{Graph Modeling the cf-mMIMO ISAC System} 


To enable GNN-based power control, we represent the ISAC system as a heterogeneous graph (HetGraph) 
$G = \langle \mathcal{V}^{\textrm{AP}}, \mathcal{V}^{\textrm{UE}}, \mathcal{V}^{\textrm{SR}}, \mathcal{E}^{\textrm{comm}}, \mathcal{E}^{\textrm{rad}} \rangle,$
where $\mathcal{V}^{\textrm{AP}}$, $\mathcal{V}^{\textrm{UE}}$, and $\mathcal{V}^{\textrm{SR}}$ denote the sets of AP, UE, and SR nodes, respectively. The sets $\mathcal{E}^{\textrm{comm}}$ and $\mathcal{E}^{\textrm{rad}}$ represent the communication and radar edges. Each AP node $m$ is initialized with its location $[x^{\textrm{AP}}_m, y^{\textrm{AP}}_m]$, sensing parameters $[q_{a_m}, q_{b_m}, q_{c_m}]$, SINR threshold $\gamma^{\textrm{thr}}_m$, and localization accuracy requirement $\nu$. The UE and SR nodes are initialized without features. Each communication edge $e^{\textrm{comm}}_{mk} \in \mathcal{E}^{\textrm{comm}}$ connects AP $m$ to UE $k$, with the channel gain $g_{mk}$ as its initial feature. Each radar edge $e^{\textrm{rad}}_{mt} \in \mathcal{E}^{\textrm{rad}}$ connects AP $m$ to SR $t$, with the radar cross-section (RCS) coefficient $\xi_{mt}$ as its initial feature.

\subsubsection{HetGNN with Node/Edge Convolutional Layers}
To capture information from all the edges and nodes in the graph, we design the three steps of the edge convolutional layer in the HetGNN as hereafter. Note that the superscript $(d)$ denotes the $(d)$-th layer, while the superscript $\textrm{X}$, and $\textrm{Y}$ represent the type of node and edge, respectively. Here, we have $\textrm{X} \in \{ \textrm{AP}, \textrm{SR}, \textrm{UE} \}$, and $\textrm{Y} \in \{ \textrm{comm}, \textrm{radar}\}$. The forward computation during each edge convolutional layer is as follows:

\textit{Aggregation:} Messages from the neighboring nodes are aggregated as follows:
\begin{align}\label{eq: aggregation}
   m(v_{i}^{\textrm{X},(d)}) = \underset{v_j \in \mathcal{A}(v_{i}^{\textrm{X}})}{\Theta^{\textrm{X},(d)}} \left( \psi^{\textrm{X},(d)} \left( v_{j}^{(d-1)}, e_{ij}^{\textrm{Y},(d)} \right) \right),
\end{align}
where \( \mathcal{A}(v_{i}^{\textrm{X}}) \) is the set of adjacent neighbors of node \( v_{i}^{\textrm{X}} \), \( v_{i}^{\textrm{X},(d)} \) is the hidden representation of node \( v_{i}^{\textrm{X}} \), \( \psi^{\textrm{X},(d)} \) and \( \Theta^{\textrm{X},(d)} \) are the MLP and pooling functions for the $\textrm{X}$ node type, respectively.\\
\textit{Node Update:} The node updates its features as:
\begin{align}\label{eq: node update}
   v_{i}^{\textrm{X},(d)} = \phi^{\textrm{X},(d)} \left( v_{i}^{\textrm{X},(d-1)}, m(v_{i}^{\textrm{X},(d)}) \right),
\end{align}
where \( \phi^{\textrm{X},(d)} \) is the node update MLP for node type $\textrm{X}$.\\
\textit{Edge Update:} The edge gets its attribute updated as:
\begin{align}\label{eq: edge update}
    e_{ij}^{\textrm{Y},(d)} = \omega^{\textrm{X},(d)} \left( v_{i}^{(d)}, e_{ij}^{\textrm{Y},(d)}, v_{j}^{(d)} \right),  \mathcal{A}(e_{ij}^{\textrm{Y}}) = \{v_{i}, v_{j} \}, 
\end{align}
where $\mathcal{A}(e_{ij}^{\textrm{Y}})$ are two end point nodes of the edge $e_{ij}^{\textrm{Y}}$, $\omega^{\textrm{Y},(d)}(\cdot)$ is the attribute update function for the edge type $\textrm{Y}$. At the last layer, say the $D$-th layer, we set the ReLu as an activation function for guaranteeing the non-negative power allocation. The AP activation is determined by the sigmoid output \( \sigma(v_{m}^{\textrm{X},(D)}) \), where the AP is active if \( \sigma(v_{m}^{\textrm{X},(D)}) \geq 0.5 \), and inactive otherwise. This threshold-based approach enables a smooth and differentiable decision process.


\subsubsection{Loss function \& model training}
To minimize the total power consumption, we define the loss functions as
\begin{align}\label{Loss fn: Sumrate}
        \mathcal{L}(\boldsymbol{\Psi}) =& \epsilon_1 \text{MSE}(\{P_{mk}\}, \{P_{mk}^{*}\}) + \epsilon_2\text{BCE}(\mathcal{M}_{\mbox{act}}, \mathcal{M}^{*}_{\mbox{act}}) \nonumber\\
        &+ \sum\limits_{m\in \mathcal{M}_{\textrm{act}}}\sum\limits_{k\in \mathcal{K}} P_k + \sum\limits_{k\in \mathcal{K}}\alpha (\gamma_{k} - \gamma^{\textrm{thr}}_{k}) \nonumber\\
        & + \beta (\bm{b}^{T} \bm{p}_{\text{sen}} -\nu \bm{p}^T_{\text{sen}} \bm{A} \bm{p}_{\text{sen}}),
\end{align}
where $\text{MSE}(\{P_{mk}\}, \{P_{mk}^{*}\})$ and $\text{BCE}(\mathcal{M}_{\mbox{act}}, \mathcal{M}^*_{\mbox{act}})$ denote the mean squared error and binary cross-entropy, respectively, comparing the model’s predictions to labels generated by the modified-BB algorithm. The coefficients $\epsilon_1$ and $\epsilon_2$ control the influence of these supervised learning terms, guiding the model while the remaining unsupervised term seeks an improved solution. The penalty factors $\alpha$ and $\beta$ are used for regulating the loss incurred, when the constraints in \eqref{P2:gamma_thr} and \eqref{P2:CRLB} are violated, respectively. 
{\color{black}
\subsubsection{Computational Complexity}
In this section, we analyze the computational complexity of the proposed HetGNN model. Following the approach in \cite{2024_SensJ_Giang_HGNN}, we divide the computational cost of GNN-based models into two main components: (i) the forward pass and (ii) the backpropagation. The complexity of each component is examined in the subsequent discussion.\\
\textbf{\textit{Forward Pass:}} The computation cost of each step in the forward pass is divided into four distinct steps as follows. For clarity, the superscript layer index $(l)$ is omitted.
\begin{itemize}
    \item \textbf{Message Generation:} We can generalize the message generation in \eqref{eq: aggregation} as $\varrho(\mathbf{X}\mathbf{W})$, where $\mathbf{W}$ represents the set of parameters of $\psi^{\textrm{X},(h)}(\cdot)$, and $\varrho(\cdot)$ is a non-linear activation function. Without loss of generality, assuming a constant embedding size $F$ across all node feature for every $D$ HetGNN layers, the computation complexity of this step is on the order of $\bigO((M+K+T) F^2 + (M+K+T))$.
    \item \textbf{Message Passing:} For each edge $e \in \mathcal{E}$, two messages of size $F$ are aggregated, resulting in a total order of $\bigO(2(MK + MT))$.
    \item \textbf{Node Embedding Update:} In \eqref{eq: node update}, applying the MLP $\phi^{\textrm{X},(h)}(\cdot)$ yields the same computational complexity as the first step of $\bigO((M+K+T) F^2 + (M+K+T))$.
    \item \textbf{Edge Embedding Update:} In \eqref{eq: edge update}, applying the MLP $\omega^{\textrm{Y},(d)}(\cdot)$ to edges yields the computational complexity on the order of $\bigO((MK + KT) F^2 + (MK + KT))$.
\end{itemize}
\textbf{\textit{Backward Pass:}}
In reverse-mode automatic differentiation, the gradient computation retraces the forward-pass graph. Therefore, the backward-pass cost matches the forward-pass cost. Hence, for an $L$-layer HetGNN with maximum initial feature dimension $F$ on the HetGraph of $K$ UEs, $M$ APs, and $T$ SRs, the total computational complexity is given by:
\begin{align} 
    \bigO(4L\left((M+K+T)F^{2} + M(K+T)\right)).
\end{align}
Compared to the model‐based algorithm in Remark~\ref{rmk:complexity}, which relies on multiple iterative loops and high computational overhead, the proposed HetGNN achieves significantly lower and more scalable complexity. After training, the model runs in time linear in the number of UEs, APs and SRs, enabling rapid and efficient deployment in large‐scale ISAC systems.
}

\section{Performance Evaluation}\label{ExperimentalEvaluation}
This section evaluates the performance of the proposed HetGNN-based framework by numerical results. 
\textcolor{black}{Regarding the proposed HetGNN, we use a three-layer model where each layer employs two MLPs with $[\text{input}-256-128]$ architecture with LeakyReLU activation for message passing and node updating. The model is trained on various generated scenarios and evaluation on unseen data.
}
For comparison, we consider the following benchmarks: 
\textcolor{black}{
\textbf{Decoupled method:} We incrementally build the active‐AP set, starting from $\vert\mathcal{M}_{\mathrm{act}}\vert=1$. For each candidate set, we solve the convex subproblem~\eqref{P2:problem} without the sensing constraint to obtain an initial power vector $\bm{\bar{p}}$. We then enforce both communication and sensing requirements via the scaling factor $\kappa = \max\left(\tfrac{\bm{b}^{\textrm{T}}\bm{\bar{p}}}{\nu \bm{\bar{p}}^{\textrm{AP}} \bm{A} \bm{\bar{p}}}, 1\right)$, where $\bm{b}$ and $\bm{A}$ are given in \eqref{eq:b_A}. If any AP exceeds its transmit power limit, the candidate set is considered infeasible. We then consider the next set or increment $|\mathcal{M}_{\mathrm{act}}|$, repeating this procedure until all constraints are satisfied.
}
\textcolor{black}{\textbf{Full AP + Constraint relaxed:} The problem in \eqref{P2:problem} is reformulated as in Section~\ref{sec: modified BB}, with all APs activated ($\mathcal{M}_{\mathrm{act}}=\mathcal{M}$), and it is then solved directly using a convex solver.
}
{\color{black}
\textbf{Deep Deterministic Policy Gradient (DDPG)}: We adopt an actor–critic framework having a pair of Q-networks and one policy network. The actor is a three‐layer MLP $(128-64-\text{output})$ with sigmoid activations; the critic is an MLP $(64-128-1)$ applied to the concatenated 1-dimensional (1D) flattened state and action.
}

\begin{table}[t] 
\centering
\color{black}
  \vspace{0.2cm}
  \renewcommand{\arraystretch}{1.5}
  \renewcommand{\tabcolsep}{0.7em}
  \caption{Key Parameters for System Simulation}
  \centering
  \resizebox{\columnwidth}{!}{\begin{tabular}{@{}lc|lc@{}}
      \toprule[1pt]\midrule[0.3pt] 
      \textbf{Parameters} &  \textbf{Value} & \textbf{Parameters} &  \textbf{Value}\\ 
      \midrule 
      Bandwidth $(B)$ &  $20$ [MHz] & Carrier Frequency $(f_c)$ &  $1900$ [MHz] \\
      Noise Power Density $(n_0)$ &  $-174$ [dBm/Hz] & Area Diameter $(D)$ &  $500$ [m]\\ 
      SINR Requirement $(\gamma^{\textrm{thr}})$ &  $-5.6$ [dB]	& Localization Accuracy $(\nu)$ &  $1$ [$\textrm{m}^2$]\\  
      Pilot Power $(\rho_{p})$ &  $100$ [mW]	& Receiver Power $(P^{\textrm{cir}})$ &  $200$ [mW]\\
      Training Duration ($\tau_{p}$) & $20$ [samples]  & Downlink Power $(\rho_{\text{DL}})$ &  $100$ [mW]\\ 
      \midrule[0.3pt] 
      \bottomrule[1pt] 
  \end{tabular}}
  \label{Table1}
  \vspace{-0.25cm}
\end{table}%

\textcolor{black}{For the network configuration in this study, all APs, SRs, and UEs are distributed uniformly at random within a circular area of diameter $D = 500 \mbox{ [m]}$, with the static target positioned at the center of that circle. All the channel gains obey the uncorrelated Rayleigh fading model. Particularly, $h \sim \mathcal{CN}(0,\mu)$, where $\mu_{n,k}$ is the large-scale fading coefficient between two devices formulated as $\mu = \textrm{FSPL} + \chi$, where $\textrm{FSPL} = -120.9 - 37.6 \log_{10}(d)\mbox{ [dB]}$ is the free-space path loss, and the shadow fading $\chi$ is generated from a log-normal Gaussian distribution with standard deviation $7 \mbox{ [dB]}$. The general system parameters are provided in Table \ref{Table1}. In this work, we train the HetGNN model on the data using multiple system settings of $K = [5,10,15,20]$ and $M = [30, 50, 80]$. }
\textcolor{black}{The supervised learning coefficients are set to $\epsilon_1=\epsilon_2=0.2$, and the unsupervised loss penalty factors to $\alpha=\beta=1$. This semi-supervised formulation balances supervised and unsupervised contributions, resulting in stable convergence and improved generalization \cite{sohn2020fixmatch}. 
}


\begin{figure}[t]

    \centering
    \begin{minipage}{0.24\textwidth}
        {\includegraphics[width=\textwidth]{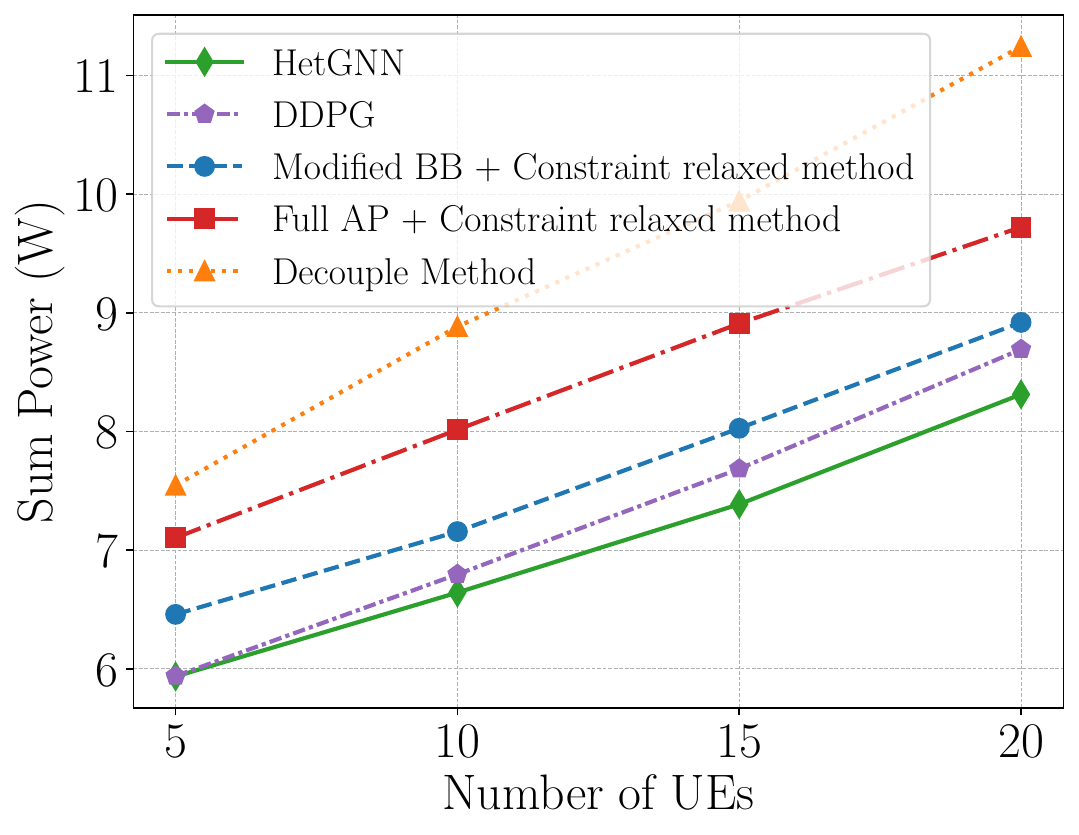}}
        \caption{Total power consumption versus the number of UEs when $M=50$.}
        \label{fig:res_sumpower}
     \end{minipage}
     \hfill
     \begin{minipage}{0.24\textwidth}
         \centering
        {\includegraphics[width=\textwidth]{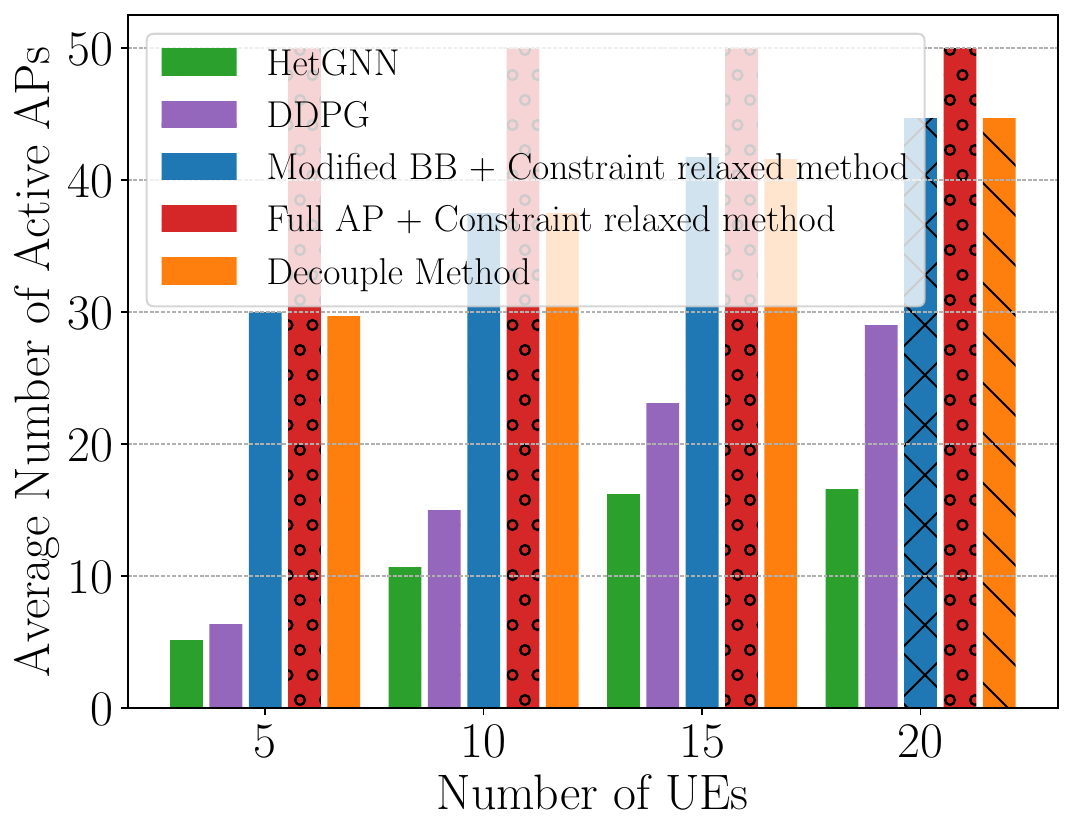}}
        \caption{Averaged number of active APs versus the number of UEs when $M=50$.}
        \label{fig:res_activeAP}
     \end{minipage}
     \vspace{-0.75cm}
\end{figure}

{\color{black}
Fig.~\ref{fig:res_sumpower} illustrates the total power consumption as a function of the number of UEs, $K \in [5,20]$, while relying on $M = 50$ available APs. The proposed HetGNN model achieves the lowest power consumption by leveraging a semi-supervised learning approach, combining insights from the modified-BB along with the relaxed method using unsupervised self-improvement. The HetGNN reduces power consumption by ~$25$\% compared to the modified-BB with the relaxed constraint method and by about $60$\% compared to when the system activates all APs. For instance, for $10$ UEs, the proposed HetGNN consumes $3$ W versus $3.8$ W for the modified-BB method. 
The DDPG model performs competitively in smaller settings ($M = 5$), matching HetGNN. However, due to the 1D flattened input structure, its performance degrades significantly as the system dimension escalates. With increasing $M$, the input dimension grows rapidly, creating a bottleneck. In larger systems, DDPG slightly outperforms the modified-BB approach but fails to match HetGNN.
Furthermore, it is important to note that a separate DDPG model is trained and evaluated for each scenario.
The fully active AP method reduces power compared to the decoupled approach, but remains inefficient due to unnecessary hardware usage from activating all APs, leading to higher overall power consumption. The decoupled method yields the highest power consumption, scaling linearly with the number of UEs due to its direct allocation strategy. At 20 UEs, it consumes around $11.4$ W, about $50$\% more than the modified-BB using relaxed constraints.

The average number of active APs is shown in Fig.~\ref{fig:res_activeAP}. The proposed HetGNN activates the fewest APs while supporting high communication and sensing performance. Specifically, it requires only $16$ APs to serve $20$ UEs, while meeting the sensing requirement, despite using less resources than the modified BB harnessing the relaxed constraint method, which activates $32$ APs. 
The DDPG model activates fewer APs than the baseline methods but still more than HetGNN, requiring $29$ APs for $20$ UEs. This reflects its limited ability to scale efficiently under increasing system size. 
By contrast, the decoupled method gradually increases the number of active APs, while searching for a feasible power allocation. As a result, it requires around $42$ APs to support $20$ UEs and the sensing task. This is almost twice the number used by HetGNN. 
}

\begin{table}[t]
\color{black}
\renewcommand{\arraystretch}{2.5} 
\setlength{\tabcolsep}{1em} 
\centering
\caption{Performance comparison between learning-based, and model-based approaches with different system scenarios.}
\label{tab:Perf_comp}
\resizebox{\columnwidth}{!}{
\begin{tabular}{@{\hspace{1em}}cccccccccc @{\hspace{1em}}}
\toprule \midrule
\multirow{2.5}{*}{\makecell{$M$}}	& \multirow{2.5}{*}{\makecell{$K$}}	& \multicolumn{4}{c}{Total Power Consumption $ \mbox{[W]}$} & \multicolumn{4}{c}{Run-time $\mbox{[s]}$}   \\ 
\cmidrule(r){3-6} \cmidrule(lr){7-10}
						&& HetGNN		& DDPG		&\makecell{Modified BB\\+ Constraint\\relaxed method}	&\makecell{Decoupled\\method}	& HetGNN		& DDPG	 &\makecell{Modified BB\\+ Constraint\\relaxed method} &\makecell{Decoupled\\method}	 \\ \hline
$30$     & $6$			& \makecell{ $4.04$ \\ $(\boldsymbol{73.46}\%)$} 	& \makecell{ $4.13$ \\ $(75.14\%)$ }	& \makecell{ $4.34$ \\  $(78.92\%)$ }	& \makecell{ $5.50 $ \\ $(\textit{100}\%)$ }	& $0.08$	& $0.07$  	& $87.10  $ 	 & $61.97$   	\\
$30$     & $10$			& \makecell{ $5.92$ \\ $(\boldsymbol{74.62}\%)$} 	& \makecell{ $6.22$ \\ $(78.37\%)$ }	& \makecell{ $6.30$ \\  $(79.45\%)$ }	& \makecell{ $7.93 $ \\ $(\textit{100}\%)$ }	& $0.08$	& $0.07$  	& $187.05   $ 	 & $131.04$  	\\
$50$     & $10$			& \makecell{ $6.64$ \\ $(\boldsymbol{74.77}\%)$} 	& \makecell{ $6.80$ \\ $(76.50\%)$ }	& \makecell{ $7.16$ \\  $(80.55\%)$ }	& \makecell{ $8.88 $ \\ $(\textit{100}\%)$ }	& $0.08$	& $0.09$  	& $931.95   $ 	 & $744.79$  	\\
$50$     & $20$			& \makecell{ $8.31$ \\ $(\boldsymbol{73.93}\%)$} 	& \makecell{ $8.69$ \\ $(77.31\%)$ }	& \makecell{ $8.92$ \\  $(79.32\%)$ }	& \makecell{ $11.24$ \\ $(\textit{100}\%)$ }	& $0.12$	& $0.14$  	& $6035.32  $ 	 & $2904.19$ 	\\
\midrule 
\bottomrule	
\end{tabular}
}
\vspace{-0.5cm}
\end{table}

{\color{black}
Table~\ref{tab:Perf_comp} compares both the total power consumption and runtime across different system settings. The proposed HetGNN consistently achieves the lowest power usage, reducing it by up to $30$\% compared to the decoupled method and $5-7$\% compared to the modified BB. For example, at $M = 30$ and $ K = 6$, HetGNN consumes only $4.04$~W, significantly lower than the decoupled method's $4.34$~W. Although the DDPG model outperforms the modified-BB method, it still consumes about $3-5$\% more energy compared to the proposed HetGNN.
In terms of runtime, HetGNN processes the CSI almost instantly ($\leq 0.12$ s), while the heuristic-based approaches require hours when $\textbf{}K=20$. 
At $M=50$, $K=20$, HetGNN and DDPG completes execution in $0.12 - 0.14$ s, compared to $6035.32$ s for the modified BB method using the relaxed constraints method, highlighting its nearly $10,000$ times faster computation. 
Unlike heuristic approaches, whose runtime escalates drastically with the number of APs, the proposed HetGNN and DDPG maintain near-constant execution time due to its lightweight arithmetic operations. However, an increase in run-time is observed in the DDPG model due to the scalability limitations of its 1D flattened input representation.
}


\section{Conclusion}\label{Conclusion}
The joint AP activation and power control problem of a cell-free ISAC network was solved by minimizing the total transmit power, while meeting demanding SINR and CRB-based sensing accuracy requirements. To solve the challenging MINLP problem, a semi-supervised HetGNN was proposed for efficiently activating the APs and power allocation efficiently. Our numerical results demonstrated that the HetGNN significantly outperforms heuristic approaches, despite reducing the power consumption and promt execution. Future work will extend this framework to multi-target and multi-user scenarios, further enhancing both scalability and efficiency.


\bibliographystyle{IEEEtran}
\bibliography{Bib1}
\end{document}